\documentclass[aip,jmp,amsmath,amssymb,preprint,]{revtex4-1}
\usepackage{amssymb,amsmath,bm}

\usepackage{color}
\usepackage{mathrsfs,amsmath}
\usepackage{mathrsfs}
\usepackage{graphicx}
\usepackage{dcolumn}
\usepackage{bm}
\usepackage{natbib}
\usepackage{amsthm}

\begin{document}
\newtheorem{thm}{Theorem}[section]
\newtheorem{lemma}[thm]{Lemma}
\newtheorem{prop}[thm]{Proposition}
\newtheorem{rem}[thm]{Remark}
\newtheorem{cor}[thm]{Corollary}

\title{Conservation-Dissipation Formalism for Soft Matter Physics: I. Equivalence with Doi's Variational Approach} 
\author{Liangrong Peng}%
\affiliation{Zhou Pei-Yuan Center for Applied Mathematics, Tsinghua University, Beijing, China, 100084}
\author{Yucheng Hu}%
\affiliation{Zhou Pei-Yuan Center for Applied Mathematics, Tsinghua University, Beijing, China, 100084}
\author{Liu Hong}
\homepage{Author to whom correspondence should be addressed. Electronic mail: zcamhl@tsinghua.edu.cn}
\affiliation{Zhou Pei-Yuan Center for Applied Mathematics, Tsinghua University, Beijing, China, 100084}
\date{\today} 

\begin{abstract}
In this paper, we proved that by choosing the proper variational function and variables, the variational approach proposed by M. Doi in soft matter physics was equivalent to the Conservation-Dissipation Formalism. To illustrate the correspondence between these two theories, several novel examples in soft matter physics, including particle diffusion in dilute solutions, polymer phase separation dynamics and nematic liquid crystal flows, were carefully examined. Based on our work, a deep connection among the generalized Gibbs relation, the second law of thermodynamics and the variational principle in non-equilibrium thermodynamics was revealed.
\end{abstract}


\maketitle

\section{Introduction}
Non-equilibrium thermodynamics was a very exciting and fruitful research field in modern physics. It not only pointed out the limitations of previous equilibrium thermodynamics and thus led us to a much broader new field beyond equilibrium, but also provided a powerful and unified framework to deal with various dissipative and irreversible processes arising in physics, chemistry, biology and so on. Due to the intrinsic complexity of non-equilibrium phenomena, there formed many ``schools'' of non-equilibrium thermodynamics during the past years. And the Conservation-Dissipation Formalism (CDF) was one of them \cite{zhu2015conservation}.

The CDF could be regarded as a regularized theory of Extended Irreversible Thermodynamics proposed by M{\"u}ller and Ruggeri \cite{muller2013extended}, Jou, Casas-V{\'a}zquez and Lebon \cite{jou1996extended}, \textit{etc}. It was mathematically rooted in the generalized nonlinear version of Onsager's reciprocal relations \cite{yong2008interesting, Peng2018Generalized} and the Godunov structure for symmetrizable hyperbolic equations \cite{godunov1961interesting, friedrichs1971systems}, which in turn guaranteed the hyperbolicity of modeling equations, regularity and globally asymptotic stability of solutions, as well as a well-behaved limit of corresponding relaxation problems, \textit{etc} \cite{zhu2015conservation}. As a rigorous formalism in mathematics, CDF has been applied to plenty of non-equilibrium systems, \textit{e.g.}, non-Fourier and non-ballistic heat conduction in nano-scales \cite{huo17}, isothermal and non-isothermal flows of compressible viscoelastic fluids\cite{zhu2015conservation, huo2016structural, Yang2018Generalized}, wave propagation in saturated porous media\cite{liu2016stability}, axonal transport with chemical reactions \cite{yan2012stability}, and so on. The authors in \cite{hong2015novel} showed an interesting connection with mesoscopic kinetic theories, like the Boltzmann equation, which put CDF on a solid foundation.

On the other hand, physicists preferred to model non-equilibrium processes through a variational approach. The adoption of variational principle had a long history in physics and mathematics. As early as 1662, Fermat introduced the principle of least time, which stated rays of light traversed the path of stationary optical length with respect to variations of the path \cite{lakshminarayanan2013lagrangian}. Later, Johann Bernoulli analyzed the famous problem of brachistochrone curves. He proposed beautiful solutions, which later led to the formal foundation of calculus of variations \cite{gelfand2000calculus}. From then on, the variational approach has become a standard tool for mathematicians and physicists. One of the most significant application would be the derivation of Lagrangian dynamics based on the least action principle \cite{goldstein2011classical}. However, the construction of a general variational approach for non-equilibrium thermodynamics was still an open problem. In the presence of friction, Rayleigh generalized the Lagrange equation by adding an extra dissipative potential as a function of velocity \cite{rayleigh1873investigation}. Recently, in the field of soft matter physics, Masao Doi borrowed Rayleigh's idea and proposed a variational function called ``Rayghleighian'' in corporation with Onsager's reciprocal relations \cite{doi2013soft, Doi2011Onsager}.

Doi's work provided new insights into the problem, but still suffered from some intrinsic limitations, like constant temperature (isothermal systems), slow kinetics without inertia effect, \textit{etc} \cite{Doi2011Onsager}. According to his derivations, the dissipative matrix (or friction coefficients) had to be symmetric. While, as stated in many recent works \cite{qian2013decomposition, Peng2018Generalized}, the anti-symmetric part of a dissipative matrix played an essential role in non-equilibrium thermodynamics as a measurement of the deviation of the non-equilibrium steady state away from the detailed balance condition. More seriously, Doi's work was restricted to irreversible processes induced by friction. As a consequence, he made the calculus of variations only with respect to velocity.

In this paper, we were going to show that, by choosing the proper variational function and variables, the variational approach proposed by Doi could give the same result as CDF. This interesting observation meant that: on the one hand, we could use CDF to extend and regularize the variational approach, which in turn kept the variational approach both mathematically correct and physically meaningful; on the other hand, since CDF admitted an equivalent variational approach, we could start the modeling of non-equilibrium processes from either way just based on our convenience. Additionally, a deep connection among the generalized Gibbs relation, the second law of thermodynamics and the variational principle could be learned from our study, which might shed lights on the development of new theories for non-equilibrium thermodynamics.

\section{General Formulation}\label{general formulation}
Unlike equilibrium thermodynamics, in non-equilibrium thermodynamics, not only classical conserved variables, like the mass, momentum and energy which were widely adopted in the formulation of continuum mechanics and hydrodynamics, but also dissipative variables related to the irreversibility of non-equilibrium processes, were required to provide a comprehensive  description. Conserved variables, as they were named, obeyed some kinds of conservation laws
which could be generally expressed as
\begin{equation}\label{con}
\frac{\partial y}{\partial t}+\nabla \cdot J=0,
\end{equation}
where the space and time coordinates $(x,t)$ were in $\Omega \times (0,\infty)$, $\Omega$ was a bounded, smooth domain in $\mathbb R^3$.
The $n$-dimensional vector function $y=y(x,t)$ represented conserved variables, $J$ was the non-equilibrium flux associated with the changes of $y$. Apparently, $J$ contained information about dissipative features of the system.
Once the form of dissipative variable $J=J(y)$ was specified, we would get a closed form of partial differential equations, based on which the dynamics of a dissipative system was completely determined.

Note that, in this paper, we focused on the derivation of macroscopic models, in which fields like mass and momentum in classical hydrodynamics were the most suitable state variables.
To generalize our formulation to mesoscopic models, distribution functions had to be used instead.
Please see, \textit{e.g.}, Refs. \cite{hong2015novel,Grmela2018Generic, Peng2018Generalized} for details.

\subsection{The Conservation-Dissipation Formalism} \label{The CDF}
Now, the central question of non-equilibrium thermodynamics became how to close the PDE system given in \eqref{con}.
To solve it, we referred to the Conservation-Dissipation Formalism \cite{zhu2015conservation}.
According to CDF, a new dissipative variable $w$, which turned to be the conjugate variable of $J$ with respect to the free energy function and would be specified later, was introduced.

The adoption of conjugate variables instead of simply taking non-equilibrium fluxes $J$ played an essential role in CDF, and had a long history in equilibrium thermodynamics (\textit{e.g.}, Legendre transformations). In the field of non-equilibrium thermodynamics, we referred to \cite{Grmela1990Hamiltonian,Grmela1998Nonlinear} and references therein. Recently, Sun \textit{et al.} \cite{sun2016nonlinear} pointed out that, by choosing the thermodynamic conjugate of an extra stress, rather than stress itself, CDF provided a suitable framework for constructing genuinely nonlinear models for non-Newtonian fluids.

The whole state variable space was given by a combination of both conserved variables and dissipative variables $(y, \nabla y, \nabla^{(2)} y, \cdots, \nabla^{(N)} y; w)$.
Here the spatial derivatives of $y$ were also included in a usual expansion of the state variable space, which was widely adopted in the mathematical modeling of complex fluids.
Meanwhile, the inclusion of spatial gradients of dissipative variables was not considered to prevent the generation of high-order PDE models.

To proceed, a strictly convex free energy (or relative entropy) function
$$
f\equiv f(y, \nabla y, \nabla^{(2)} y, \cdots, \nabla^{(N)} y; w)
$$
was further specified to characterize the system dissipation.
As we claimed, $w$ and $J$ were conjugate variables with respect to $f$, which meant $J=\frac{\partial}{\partial w}  f(y, \nabla y, \cdots, \nabla^{(N)} y; w)$. The time evolution equation of the free energy $f$ was given by the generalized Gibbs relation,
\begin{equation*}
\begin{aligned}
\frac{\partial f}{\partial t}
=&\frac{\partial f}{\partial y}\frac{\partial y}{\partial t}
+\sum_{i=1}^{N} \frac{\partial f}{\partial (\nabla^{(i)} y)}\frac{\partial \nabla^{(i)} y}{\partial t}
+\frac{\partial f}{\partial w}\frac{\partial w}{\partial t}
\\
=& \nabla \cdot
\bigg\{ \sum_{i=1}^{N}\sum_{j=1}^{i} (-1)^{j-1} \bigg [\nabla^{(j-1)} \cdot \frac{\partial f}{\partial (\nabla^{(i)} y)}\bigg ] \cdot \frac{\partial \nabla^{(i-j)} y}{\partial t}
\bigg\}\\
&
+\bigg\{ \sum_{i=1}^N (-1)^i \bigg [ \nabla^{(i)} \cdot \frac{\partial f}{\partial (\nabla^{(i)} y)} \bigg] + \frac{\partial f}{\partial y}\bigg\}
 \frac{\partial y}{\partial t}
+\frac{\partial f}{\partial w}\frac{\partial w}{\partial t} \\
=& \nabla \cdot
\bigg\{ \sum_{i=1}^{N}\sum_{j=1}^{i} (-1)^{j-1} \bigg [\nabla^{(j-1)} \cdot \frac{\partial f}{\partial (\nabla^{(i)} y)} \bigg ] \cdot \frac{\partial \nabla^{(i-j)} y}{\partial t}
\bigg\}
- \frac{\delta f}{\delta y} (\nabla \cdot J) +J \frac{\partial w}{\partial t}
\\
=& \nabla \cdot
\bigg\{ \sum_{i=1}^{N}\sum_{j=1}^{i} (-1)^{j-1} \bigg [\nabla^{(j-1)} \cdot \frac{\partial f}{\partial (\nabla^{(i)} y)} \bigg ] \cdot \frac{\partial \nabla^{(i-j)} y}{\partial t} - J \frac{\delta f}{\delta y}
\bigg\}
+J \cdot (\frac{\partial w}{\partial t} + \nabla \frac{\delta f}{\delta y} ) \\
\equiv & \nabla \cdot J^f -\sigma^f,
\end{aligned}
\end{equation*}
in which
$J^f=\{ \sum_{i=1}^{N}\sum_{j=1}^{i} (-1)^{j-1} [\nabla^{(j-1)} \cdot \frac{\partial f}{\partial (\nabla^{(i)} y)}] \cdot \frac{\partial \nabla^{(i-j)}  y}{\partial t} - J \frac{\delta f}{\delta y} \}$
and $\sigma^f=-J\cdot(\frac{\partial w}{\partial t} + \nabla \frac{\delta f}{\delta y})$ denoted entropy flux and entropy production rate respectively.
And the functional derivative $\frac{\delta f}{\delta y}$ was defined as
\begin{equation*}
\frac{\delta f}{\delta y}
=\frac{\partial f}{\partial y}+\sum_{i=1}^N (-1)^i\nabla^{(i)} \cdot \frac{\partial f}{\partial (\nabla^{(i)} y)}.
\end{equation*}
In the first example in Section \ref{example1}, we had the free energy $f=f(y,w)$. Then $\frac{\delta f}{\delta y}$ reduced to standard partial derivative $\frac{\partial f}{\partial y}$; while in the last two cases \ref{example2} and \ref{example3}, $f=f(y,\nabla y; w)$, therefore we had $\frac{\delta f}{\delta y}=\frac{\partial f}{\partial y} -\nabla \cdot \frac{\partial f}{\partial \nabla y}$.

Note that there was a minus sign in front of $\sigma^f$, since here we adopted the free energy function instead of entropy. In order to keep $\sigma^f\geq0$ in accordance with the second law of thermodynamics, we referred to the generalized Onsager's reciprocal relations \cite{onsager1931reciprocal, zhu2015conservation, Peng2018Generalized} between non-equilibrium forces and fluxes as
\begin{equation}\label{w1}
-(\frac{\partial w}{\partial t}+\nabla \frac{\delta f}{\delta y})=M^{-1}\cdot J,
\end{equation}
where $M \equiv M(y, \nabla y, \cdots, \nabla^{(N)} y; w)$ was called the dissipation matrix and was strictly positive definite.

Now Eqs. \eqref{con} and \eqref{w1} together composed a closed PDE system in the form of
\begin{equation}\label{cdf}
\partial_t U + \mathop\sum^3_{j=1}\partial_{x_j}F_j(U)=Q(U)
\end{equation}
with
\begin{equation*}
    U=
    \begin{pmatrix}
    y\\w
    \end{pmatrix}
,  \quad
      \mathop\sum^3_{j=1}\partial_{x_j}F_j(U)=\nabla\cdot
      \begin{pmatrix}
      \frac{\partial f}{\partial w}\\ \frac{\delta f}{\delta y} I
      \end{pmatrix},
\end{equation*}
\begin{equation*}
   Q(U)
   =\tilde M \cdot \frac{\delta f}{\delta U}
   =   \begin{pmatrix}
   &0 & 0 \\
   &0 & -M^{-1}
   \end{pmatrix}
   \cdot
   \begin{pmatrix}
      \frac{\delta f}{\delta y} \\
      \frac{\delta f}{\delta w}
   \end{pmatrix}
   =
   \begin{pmatrix}
   0\\-M^{-1}\cdot \frac{\partial f}{\partial w}
   \end{pmatrix}.
\end{equation*}
On the left-hand side of Eq. \eqref{cdf}, $y$ denoted conserved variables, such as mass, momentum, and total energy, while $w$ denoted dissipative variables.
$I$ was the identity matrix,
$\frac{\partial f}{\partial w}=J$ and $\frac{\delta f}{\delta y} I$ represented fluxes corresponding to conserved and dissipative variables $(y, w)$, respectively.
On the right-hand side of Eq. \eqref{cdf}, $(-M^{-1}\cdot \frac{\partial f}{\partial w})$ was the nonzero source, which vanished at equilibrium.
It was worthy to emphasize that, the dissipation matrix $M$ was strictly positive definite, rather than semi-positive \cite{yong2008interesting}.
From a physical point of view, the system \eqref{cdf} reached the steady state if and only if the minimum of free energy was attained with respect to dissipative variables \cite{zhu2015conservation}.

\subsection{The variational approach}
In the last section, we have closed the evolution system \eqref{con} by deriving thermodynamically admissible constitutive relations with CDF.
In this section, we were going to show that, by choosing the proper variational function and variables as suggested by CDF, the variational approach proposed by M. Doi in soft matter physics \cite{doi2013soft, Doi2011Onsager} would lead to exactly the same results obtained by CDF.

The original version of Doi's variational principle was based on phenomenological equations, which essentially showed that the time evolution of a physical system was determined by the balance of a potential force and a generalized frictional force.
From the view of physics, the potential force drove the system into a state of potential minimum, while the frictional force resisted the trend. It was shown that this variational principle was valid for many problems in soft matter physics \cite{doi2013soft}, and more recently was applied to formulate hydrodynamics of thin films \cite{Xu2015A} and viscoelastic filaments \cite{Zhou2018Dynamics} and solid toroidal islands \cite{Jiang2019Application}, to construct boundary conditions for liquid-vapor flows and immiscible two-phase flows \cite{Xu2017Hydrodynamic}, to explain the deposition patterns of two droplets next to each other \cite{Hu2017Deposition}.

To see the result, we followed Doi's original derivation by introducing a total Rayleighian function $R=\dot{A}+\Phi$, which consisted of two physically different terms. The first term $\dot{A}=d A/d t$ represented the rate of total free energy change, while the second term $\Phi$ was called the dissipation function. Notice that $\Phi$ was the half of entropy production rate of the system. In accordance with notations used in CDF in the last section, we specified
\begin{eqnarray}
A&=&\int_{\Omega} f(y, \nabla y, \cdots, \nabla^{(N)} y; w)dx,\\
\Phi&=&\frac{1}{2} \int_{\Omega}  J^T \cdot M^{-1} \cdot J dx,
\end{eqnarray}
where $J=J(y, \nabla y, \cdots, \nabla^{(N)} y; w)$ and $M=M(y, \nabla y, \cdots, \nabla^{(N)} y; w)$. Consequently, by the Reynold's transport theorem and generalized Gibbs relation, we had
\begin{equation*}
\begin{aligned}
\dot{A}
&=\int_{\Omega} (\frac{\delta f}{\delta y} \frac{\partial y}{\partial t} + \frac{\partial f}{\partial w} \frac{\partial w}{\partial t})dx
=\int_{\Omega} (-\frac{\delta f}{\delta y} \nabla \cdot J + \frac{\partial f}{\partial w}\frac{\partial w}{\partial t})dx
=\int_{\Omega} J \cdot(\frac{\partial w}{\partial t}+\nabla \frac{\delta f}{\delta y})dx,
\end{aligned}
\end{equation*}
by assuming the surface integral $\int_{\Omega} \nabla \cdot (J^f + v f) dx =\int_{\partial \Omega} (J^f + v f) \cdot d S=0$ vanished at boundary ${\partial \Omega}$.

According to the variational principle, time evolution of a given dissipative system could be totally specified by minimizing the Rayleighian function $R$ with respect to the dissipative variable $J$, \textit{i.e.},
\begin{equation}
\frac {\delta R}{\delta J}=\frac {\delta \dot{A}}{\delta J}+\frac{\delta \Phi}{\delta J}=0.\label{DOI_GEN_EQU}
\end{equation}
Note, in Doi's work, the authors minimized the Rayleighian function $R$ with respect to the velocity $v$ instead of $J$, since the original derivation was generally restricted to irreversibility caused by friction.

Inserting formulas of $\dot{A}$ and $\Phi$ into Eq. \eqref{DOI_GEN_EQU}, we could deduce
\begin{equation*}
\frac{\partial w}{\partial t} +\nabla \frac{\delta f}{\delta y} = -M^{-1}\cdot J,
\end{equation*}
which was exactly the same relation obtained by CDF in Eq. \eqref{w1}. In this sense, Doi's variational approach was consistent with CDF. Especially, if the free energy $f$ only depended on conserved variables $y$ and its spatial derivatives ($f=f(y, \nabla y, \cdots, \nabla^{(N)} y)$), the variational approach would lead to
\begin{equation*}
\nabla \frac{\delta f}{\delta y}=-M^{-1}\cdot J.
\end{equation*}
The same conclusion could be attained by CDF too.

Now it was seen that, with the help of CDF, the new version of variational approach overcame most of its former limitations. It was no longer restricted to friction induced irreversibility. Effects of inertia and non-equilibrium temperature would be readily included into the modeling. The dissipation matrix could depend on state variables and have an anti-symmetric part too.

\subsection{Physical insights}
Generally speaking, to describe the time evolution of a given irreversible process, the macroscopic or mesoscopic models should consist of a mechanical part and a thermodynamic part \cite{Peshkov2017Continuum, Grmela2017Hamiltonian}. The mechanics, such as the Hamiltonian equations in classical mechanics, was directly related to conservation laws, which was time reversible, entropy-preserving and non-dissipative; while the thermodynamics emerged as a consequence of statistical averaging of microscopic freedoms in a macroscopic (or mesoscopic) description, and was characterized by entropy functions. It was time irreversible, generalized gradient and dissipative \cite{Grmela2018Generic}. A unification of the mechanical part and the thermodynamic part served as the core of a successful non-equilibrium theory.

In CDF, these two parts were properly combined into one PDE system. The left-hand side of Eq. \eqref{cdf} represented the reversible continuum mechanics in the form of local conservation laws. Recall that both Hamiltonian mechanics and classical hydrodynamics, like the Euler and Navier-Stokes equations, could be casted into it. Meanwhile, the right-hand side of Eq. \eqref{cdf} were rewritten into an abstract compact form $\tilde M \cdot \frac{\delta f}{\delta U}$. Here $\frac{\delta f}{\delta U}$ was known as non-equilibrium forces raised by entropy production, and $\tilde M$ was Onsager's coefficient matrix linking non-equilibrium forces and fluxes. In general, it was a semi-positive definite matrix, with degenerate zero eigenvalues corresponding to the conservation of mass, momentum, total energy and so on. At the same time, the second law of thermodynamics was preserved through the famous entropy condition for symmetrizable hyperbolic systems of first-order PDEs \cite{Godunov1961An,friedrichs1971systems,Yong2004Entropy,zhu2015conservation}, due to which Eq. \eqref{cdf} could be casted into the Godunov structure, a form of gradient dynamics guaranteeing the growth of entropy and consequently the approach to equilibrium \cite{Grmela2017Hamiltonian}.

The reversibility and irreversibility of time evolutionary dynamics in non-equilibrium thermodynamics were extensively studied within the framework of GENERIC \cite{Grmela1997dynamics, Ottinger1997dynamics}. According to GENERIC, the Poisson bracket corresponded to reversible mechanics, while the dissipative bracket generated irreversible thermodynamics. The first and second laws of thermodynamics were guaranteed simultaneously by degeneracy requirements. Mathematically, GENERIC was a direct extension of the Hamiltonian equations and Ginzburg-Landau equations, and was closely related to some version of CDF \cite{Grmela2017Hamiltonian}.

It was well known that the Hamiltonian dynamics for reversible processes allowed a variational formulation -- principle of least action defined through the Lagrangian function. However, such formulation did not readily extend to irreversible processes. Interestingly, with the help of contact geometry, GENERIC allowed a true variational principle -- the total entropy generated during the time evolution reached its extremum \cite{Grmela2018Generic}. And Doi's variational approach discussed above could be considered as a special case of it, in which the Rayleighian $R$, interpreted as the action functional of physical systems \cite{onsager1931reciprocal, doi2013soft, Xu2015A, Zhou2018Dynamics, Jiang2019Application, Xu2017Hydrodynamic, Hu2017Deposition}, reached its extremum with respect to dissipative fluxes $J$, $\delta R/ \delta J=0$. This conclusion in some sense clarified the physical meanings of CDF.

\section{Applications}

In this section, we were going to explore several novel examples in soft matter physics to further illustrate the correspondence between CDF and the variational approach.

\subsection{Particle diffusion in dilute solutions} \label{example1}
As a first application, we considered the diffusion of Brownian particles in dilute solutions. The particle density $n(x, t)$ satisfied the conservation law of mass, \textit{i.e.},
\begin{equation}\label{diffcon}
\frac{\partial n}{\partial t}+\nabla \cdot (nv) =0,
\end{equation}
where $v(x, t)$ was the average velocity of particles. Apparently, particle density $n$ was a conserved variable, while velocity $v$ was dissipative due to the existence of friction.

For this system, we specified a free energy function as
\begin{equation}
f=\frac{1}{2}nv^2+nU(x)+k_BTn\ln n,
\end{equation}
where $\frac{1}{2} n v^2$ was the kinetic energy, $U(x)$ represented the potential energy of a single particle due to the presence of external force  fields (\textit{e.g.}, the gravitation), and $-k_B n\ln n$ was the entropy for particle mixing with constant temperature $T$.
With respect to the free energy, it was easy to verify that the conjugate variable of flux $J=n v$ was ${\partial f}/{\partial J}=v$. Thus we could choose the particle velocity $v$ as a dissipative variable in CDF. The time changes of the free energy followed the generalized Gibbs relation,
\begin{equation*} \label{EIT-GEN-EQU}
\begin{aligned}
\frac{\partial {}}{\partial t} f(n, v)
&=\frac{\partial f}{\partial n} \frac{\partial n}{\partial t} + \frac{\partial f}{\partial v} \cdot \frac{\partial v}{\partial t}\\
&=[\frac{1}{2}v^2 +U(x)+k_BT\ln n+k_BT] \cdot [-\nabla\cdot(nv)]+nv \cdot \frac{\partial v}{\partial t}\\
&=-\nabla\cdot\big\{nv\big[\frac{1}{2}v^2 +U(x)+k_B T\ln n+k_B T \big]\big\} + nv\cdot[\nabla U + k_B T \frac{\nabla n}{n} + \frac{\partial v}{\partial t} + v\cdot\nabla v]\\
&\equiv \nabla \cdot J^f - \sigma^f.
\end{aligned}
\end{equation*}
Here the entropy flux was given by $J^f=-nv(\frac{1}{2}v^2 +U+k_BT\ln n+k_BT)$, and the entropy production rate was $\sigma^f=-nv\cdot(\nabla U+k_B T \frac{\nabla n}{n} + \frac{\partial v}{\partial t} +v\cdot\nabla v) \geq 0$.

It was recognized that $J=n v$ was the non-equilibrium flux and $-(\nabla U+k_B T \frac{\nabla n}{n} + \frac{\partial v}{\partial t} +v\cdot\nabla v)$ was the corresponding non-equilibrium force. Especially, if we chose $M=n/\zeta$ in accordance with the Onsager's relation, where $\zeta>0$ was the friction coefficient, we arrived at the constitutive relation
\begin{equation*}
v=-\frac{1}{n\zeta}(n\nabla U+k_BT\nabla n+n \frac{\partial v}{\partial t} + nv\cdot\nabla v),
\end{equation*}
or
\begin{equation}\label{diffusion-momentum}
\frac{\partial}{\partial t} (nv)+\nabla\cdot (n v v)=-n\nabla U-k_BT\nabla n-n\zeta v,
\end{equation}
by using the continuity equation.
Above equation turned to be the classical momentum equation for particle motion by considering the external potential force $n \nabla U$, friction force $n\zeta v$, as well as entropic force $k_BT\nabla n$ arising from  particle mixing.

In addition, if the free energy function was assumed not to rely on the particle velocity, \textit{i.e.}, $f(n)=nU(x)+k_B Tn\ln n$, then by repeating the same procedure above, we obtained
\begin{equation*}
\begin{aligned}\label{dilute DOI EIT}
\frac{\partial }{\partial t}f(n)
&=\frac{\partial f}{\partial n} \frac{\partial n}{\partial t}
=-\nabla\cdot [nv(U+k_B T\ln n+k_B T)]+nv\cdot(\nabla U + k_BT \frac{\nabla n}{n} ),
\end{aligned}
\end{equation*}
which meant the entropy flux $J^f=-nv(U+k_BT\ln n+k_BT)$ and the entropy production rate $\sigma^f=-nv\cdot(\nabla U+k_BT \frac{\nabla n}{n} )$. Consequently, the constitutive relation became
\begin{equation}\label{diffusion-overdamped}
v=-\frac{1}{n\zeta}(n\nabla U+k_BT\nabla n).
\end{equation}

To construct an equivalent variational approach, we set two parts of the Rayleighian as
\begin{equation*}
A=\int f(n, v)dx=\int \bigg[\frac{1}{2}nv^2+nU(x)+k_BTn\ln n\bigg]dx,
\end{equation*}
and
\begin{equation*}
\Phi= \frac{1}{2} \int  J^T \cdot  M^{-1} \cdot J dx=\frac{1}{2}\int \frac{\zeta(nv)^2}{n} dx.
\end{equation*}
Now a key step was to calculate the time derivative of the total free energy,
\begin{equation*}
\begin{aligned}
\dot{A}&=\int nv\cdot \frac{\partial v}{\partial t}+(\frac{1}{2}v^2+ U +k_BT+k_BT\ln n)\frac{\partial n}{\partial t} dx\\
&=\int nv\cdot \frac{\partial v}{\partial t}-(\frac{1}{2}v^2+ U +k_BT+k_BT\ln n)\nabla\cdot(nv)dx\\
&=\int nv\cdot(\frac{\partial v}{\partial t}+v\cdot\nabla v+\nabla U+ k_B T \frac{\nabla n}{n} )dx, \\
\end{aligned}
\end{equation*}
where flux $J=n v$ was assumed to be vanished at the boundary. Substituting above formulas into Eq. \eqref{DOI_GEN_EQU}, we arrived at the same result as Eq. \eqref{diffusion-momentum}.

In what follows, we adopted an alternative way to derive Eq. \eqref{diffusion-overdamped} from \eqref{diffusion-momentum}. We considered the over-damped  limit when the friction coefficient $\zeta \rightarrow \infty$. By applying the Maxwell iteration \cite{yong2004diffusive}, we could deduce that
\begin{equation*}
\begin{aligned}
v&=-\frac{1}{n\zeta}(n\nabla U+k_BT\nabla n+n\frac{\partial v}{\partial t}+nv\cdot\nabla v),\\
&=-\frac{1}{n\zeta}(n\nabla U+k_BT\nabla n)+\frac{1}{\zeta^2}\frac{\partial}{\partial t}(\nabla U+ k_B T \frac{\nabla n}{n} )+o(\zeta^{-2}),\\
&=-\frac{1}{n\zeta}(n\nabla U+k_BT\nabla n)+o(\zeta^{-1}).
\end{aligned}
\end{equation*}
The leading term gave the desired result.


\subsection{Phase separation in polymeric solutions}\label{example2}
Next, we considered the phenomenon of phase separation emerging in polymer solutions. Its variational formulation has been illustrated by Zhou, Zhang and E \cite{zhou2006modified}, so here we only focused on the derivation based on CDF.

Let $v_p(x, t)$ and $v_s(x, t)$ be average velocities of polymers and solvent molecules at point $x$ and time $t$ respectively. Then volume fractions of polymers $\phi(x, t)$ and solvent molecules $1-\phi(x, t)$ satisfied following continuity equations,
\begin{align}
  &\frac{\partial \phi}{\partial t}=-\nabla \cdot (\phi v_p ),\label{0.1} \\
  &\frac{\partial (1-\phi)}{\partial t}=-\nabla \cdot [(1-\phi)v_s] \label{0.2}.
\end{align}
Introduce the volume-averaged velocity of solutions as $v=\phi v_p+(1-\phi)v_s$. Then the summation of Eqs. \eqref{0.1} and \eqref{0.2} led to the incompressible condition
\begin{equation}\label{0.3}
  \nabla \cdot v=0.
\end{equation}
Furthermore, $v$ obeyed the conservation law of total momentum,
\begin{equation}\label{0.4}
   \frac{d v}{d t} \equiv
   \frac{\partial v}{\partial t}+v\cdot \nabla v
   =-\nabla p+\nabla \cdot \tau_e +\nabla \cdot \tau_v,
\end{equation}
where $p$ was the thermodynamic pressure, $\tau_e$ and $\tau_v$ were symmetric tensors and denoted the elastic stress and viscous stress, respectively.

The specific entropy of the solution was constituted by three parts: the entropy for solution mixing, the entropy for phase separation \cite{cahn1958free} and the conformational entropy of polymer chains, \textit{i.e.},
\begin{equation}\label{2}
s(\phi, \nabla \phi, b)=-\eta(\phi)-\frac{1}{2}\alpha_0|\nabla \phi|^2-\frac{1}{2}b^2,
\end{equation}
where $\alpha_0\geq 0$ was a positive constant, $b I$ was the bulk stress tensor arising from polymer configurations. The mixing entropy could be modeled by the classical Flory-Huggins theory \cite{doi2013soft},
\begin{equation}\label{3}
\eta(\phi)
 =\frac{1}{m_p}\phi \ln \phi + \frac{1}{m_s}(1-\phi)\ln(1-\phi)+\chi \phi(1-\phi),
\end{equation}
where $m_p$ and $m_s$ denoted molecular weights of polymers and solvent molecules separately. $\chi$ was the effective Flory interaction parameter.

The specific internal energy included the kinetic energy of solutions and elastic energy of polymers,
\begin{equation}\label{0.5}
  u(v, \tau_s)=\frac{1}{2}|v|^2+\frac{1}{2}tr(\tau_s),
\end{equation}
where $\tau_s=\tau_e-\frac{\partial s}{\partial \nabla \phi}\otimes \nabla \phi=\tau_e+\alpha_0\nabla \phi\otimes \nabla \phi$
was a symmetric tensor and was recognized as the shear stress.
The symbol $\otimes$ represented the tensor product, $(\nabla \phi\otimes \nabla \phi)_{ij}=\frac{\partial \phi}{\partial {x_i}}  \frac{\partial \phi}{\partial {x_j}}$.
Notice that the elastic energy of polymers was non-negative ($tr (\tau_s)\geq 0$) in accordance with the Hookean-dumbbell models \cite{doi1988theory}. Consequently, the free energy function became
\begin{equation}\label{0.6}
f=u-Ts=\eta(\phi)+\frac{1}{2}\alpha_0|\nabla \phi|^2  + \frac{1}{2}b^2+ \frac{1}{2}|v|^2+\frac{1}{2}tr(\tau_s),
\end{equation}
where the temperature $T$ was assumed to be $1$ for an isothermal process.

Now, we could firstly use the generalized Gibbs relation to calculate the time evolution of the entropy $s(\phi, \nabla \phi, b)$ as
\begin{align*}
\frac{ds}{dt}=& -\frac{\partial \eta}{\partial \phi}\frac{d\phi}{dt}-\alpha_0\nabla \phi \cdot \frac{d\nabla \phi}{dt}- b\frac{d b}{dt}\\
    =&-\frac{\partial \eta}{\partial \phi}\frac{d\phi}{dt}-\alpha_0\nabla \phi \cdot \bigg(\nabla {\frac{d\phi}{dt}} - \nabla v\cdot \nabla \phi\bigg)-b\frac{d b}{dt}\\
    =&-\nabla \cdot \bigg(\alpha_0\nabla \phi \frac{d\phi}{dt}\bigg)
    -\bigg( \frac{\partial \eta}{\partial \phi}- \alpha_0\Delta \phi \bigg) \frac{d\phi}{dt}+ \alpha_0\nabla \phi \cdot (\nabla v\cdot \nabla \phi)- b\frac{d b}{dt}.
\end{align*}
Then, the time evolution of the free energy $f(\phi, \nabla \phi, v, b, \tau_s)$ was given by
\begin{align*}
    \frac{df}{dt}=&v\cdot \frac{dv}{dt}+\frac{1}{2}\frac{d}{dt}tr(\tau_s)-\frac{ds}{dt}\\
    =&v\cdot (-\nabla p+\nabla \cdot \tau_e +\nabla \cdot \tau_v )+\frac{1}{2} tr\bigg(\frac{D}{Dt}\tau_s \bigg)+\nabla v: \tau_s +\nabla \cdot \bigg(\alpha_0\nabla \phi \frac{d\phi}{dt}\bigg)\\
    &-\bigg( \frac{\partial \eta}{\partial \phi}- \alpha_0\Delta \phi \bigg ) \nabla\cdot[\phi(1-\phi)(v_p-v_s)]- \alpha_0\nabla \phi \cdot (\nabla v\cdot \nabla \phi)+ b\frac{d b}{dt}\\
    =& \nabla \cdot \bigg[\alpha_0\nabla\phi\frac{d\phi}{dt} -\bigg( \frac{\partial \eta}{\partial \phi}- \alpha_0\Delta \phi \bigg)\phi(1-\phi)(v_p-v_s)
    +v\cdot (-pI +\tau_e +\tau_v)\bigg]\\
    &+\frac{1}{2} tr\bigg(\frac{D}{Dt}\tau_s \bigg)+  [\phi(1-\phi)(v_p-v_s)] \cdot \nabla \bigg( \frac{\partial \eta}{\partial \phi}- \alpha_0\Delta \phi \bigg )\\
    & -\nabla v : \big(\alpha_0\nabla \phi\otimes\nabla \phi +\tau_e-\tau_s+\tau_v \big) + b\frac{d b}{dt}\\
    =& \nabla \cdot J^f+[\phi(1-\phi)(v_p-v_s)] \cdot \nabla \bigg( \frac{\partial \eta}{\partial \phi}- \alpha_0\Delta \phi-\alpha_1 b \bigg )- \nabla v : \tau_v\\
    &+\frac{1}{2} tr\bigg[\frac{D}{Dt}\tau_s-\alpha_2 (\nabla v + (\nabla v)^T)\bigg]+  b\bigg \{ \frac{d b}{dt}-\alpha_1 \nabla \cdot [\phi(1-\phi)(v_p-v_s)] \bigg\},
\end{align*}
where the entropy flux $J^f=\alpha_0\nabla\phi\frac{d\phi}{dt} -\big( \frac{\partial \eta}{\partial \phi}- \alpha_0\Delta \phi-\alpha_1 b \big )\phi(1-\phi)(v_p-v_s)+v\cdot (-pI +\tau_e +\tau_v)$. $\frac{D}{Dt} \tau_s=\frac{d}{dt}\tau_s-(\nabla v)^T \cdot \tau_s-\tau_s \cdot \nabla v$ denoted the upper-convected time derivative.
Notice that we utilized the material derivative of the free energy $df/dt$ for notational convenience in Section \ref{example2} and \ref{example3}.
Since $df/dt=\partial f/ \partial t + \nabla \cdot (v f) $ for incompressible fluids, it fitted into the framework of general formulation in Section \ref{general formulation}.
During above derivation, we have used continuity equation $d\phi/dt=-\nabla[\phi(1-\phi)(v_p-v_s)]$ and identity $\frac{1}{2} \frac{d}{dt}tr(\tau_s)= \frac{1}{2} tr(\frac{D}{Dt}\tau_s )+ \nabla v : \tau_s$ in the second step. The colon $:$ stood for the double inner product between two second-order tensors, \textit{i.e.}, $A:B=\sum_{i,j}A_{ij}B_{ij}$. While in the last step, without affecting entropy production rate, two additional parameters $\alpha_1$ and $\alpha_2$ were introduced, accounting for effects of velocity difference on polymer compressibility and solution velocity gradient on the shear stress, respectively.

To guarantee the non-negativeness of
\begin{align*}
\sigma^f=&-[\phi(1-\phi)(v_p-v_s)] \cdot \nabla \bigg( \frac{\partial \eta}{\partial \phi}- \alpha_0\Delta \phi-\alpha_1 b \bigg)+ \nabla v : \tau_v\\
    &-\frac{1}{2} tr\big[\frac{D}{Dt}\tau_s-\alpha_2 (\nabla v + (\nabla v)^T)\big]-  b\bigg \{ \frac{d b}{dt}-\alpha_1 \nabla \cdot [\phi(1-\phi)(v_p-v_s)] \bigg\},
\end{align*}
CDF suggested following constitutive relations,
\begin{align*}
&v_p -v_s=-M(\phi) \nabla \bigg( \frac{\partial \eta}{\partial \phi}- \alpha_0\Delta \phi-\alpha_1 b \bigg), \\
&\tau_v= \zeta[\nabla v +(\nabla v)^T],\\
&\frac{d b}{dt}- \alpha_1 \nabla \cdot [\phi(1-\phi)(v_p-v_s)] = -\frac{1}{\epsilon} b, \\
&\frac{d}{dt}\tau_s-(\nabla v)^T \cdot \tau_s-\tau_s \cdot \nabla v -\alpha_2 [\nabla v + (\nabla v)^T] = -\frac{1}{\xi} \tau_s,
\end{align*}
The first relation represented the fact that the velocity difference between polymers and solvent molecules was caused by chemical potentials from mixing, phase separation and polymer configuration, separately. $M(\phi) \geq 0$ was a coefficient depending on the volume fraction of polymers $\phi$. The second formula was the Newton's law of viscosity with $\zeta \geq 0$. The third and fourth relations both belonged to relaxation equations with $\epsilon, \xi>0$ representing typical relaxation times for polymer compressing and solution shearing, respectively. In particular, the last equation was the upper-convected Maxwell model.

Finally, by using CDF, we arrived at the same governing equations for phase separation in polymer solutions, which has been studied by Zhou, Zhang and E
based on the variational approach \cite{zhou2006modified}, \textit{i.e.},
\begin{equation}
\begin{aligned}
  &\frac{\partial \phi}{\partial t}+v\cdot \nabla \phi=\nabla \cdot g, \label{6.1}\\
  &\frac{\partial v}{\partial t}+v\cdot \nabla v=-\nabla p+\nabla \cdot \tau_s- (2\alpha_0 \Delta \phi)\nabla \phi +\zeta \Delta v,\\
  &\frac{\partial b}{\partial t}+v\cdot \nabla b= -\frac{1}{\epsilon} b  - \alpha_1 \nabla \cdot g,\\
  &\frac{\partial \tau_s}{\partial t}+v\cdot \nabla \tau_s-(\nabla v)^T \cdot \tau_s-\tau_s \cdot \nabla v= -\frac{1}{\xi} \tau_s +\alpha_2 [\nabla v + (\nabla v)^T],\\
  &\nabla\cdot v=0,\\
\end{aligned}
\end{equation}
where $g=\phi(1-\phi)M(\phi) \nabla \big( \frac{\partial \eta}{\partial \phi}- \alpha_0\Delta \phi-\alpha_1 b \big)$ was the osmotic pressure and was slightly different from the one $\phi(1-\phi)M(\phi) \nabla \big( \frac{\partial \eta}{\partial \phi}- \alpha_0\Delta \phi\big)+M(\phi)\nabla(\alpha_1 b)$ defined in Ref. \cite{zhou2006modified}

\subsection{Flows of liquid crystals in nematic phase}\label{example3}
In this section, we were going to discuss the continuum theory of liquid crystals in the nematic phase, which was an intermediate material between solids and fluids.
The conservation laws and constitutive equations of nematic liquid crystals were developed by Ericksen \cite{Ericksen1961Conservation} and Lesile \cite{Leslie1979Theory} in the 1960's.
Later, Lin and Liu \cite{Lin1995Nonparabolic, Lin2000Existence} simplified the Ericksen-Lesile (E-L) model by introducing a penalty approximation of the optical director, and reducing the bulk energy density (Oseen-Frank energy) into two terms.
The simplified model turned out to retain most mathematical properties of interest of the E-L theory \cite{Lin1995Nonparabolic, Lin2000Existence}.

For simplicity, we restricted ourself to isothermal situations.
The nematic liquid crystal was usually treated as incompressible materials and its velocity field of flows was denoted as $v \in \mathbb R^3$.
To characterize the orientational preference of rod-like molecules of liquid crystals, a direction vector $d\in \mathbb R^3$ was introduced.
Consequently, the conservation laws of the mass, momentum and angular momentum became
\begin{equation}\label{liquid-1}
\begin{aligned}
&\nabla \cdot v=0,\\
&\frac{\partial v}{\partial t}+(v \cdot \nabla v)= \nabla \cdot \tau,\\	
&\frac{\partial d}{\partial t}+(v \cdot \nabla d) =q.
\end{aligned}
\end{equation}
Here $q \in \mathbb R^3$ denoted the force moment, $\tau=-p I+ \tau_v + \tau_e$ was the stress tensor and included three different contributions: the isotropic thermodynamic pressure $p$, the viscous stress $\tau_v(v)$ and elastic stress $\tau_e(d)$.
To close above equations, the constitutive equations for $\tau$ and $q$ were needed.
A variational approach for modeling nematic liquid crystal flows was proposed by Liu and Sun \cite{Liu2009On}, again we focused on CDF.

The free energy function for this system was specified as
\begin{equation} \label{liquid crystal free energy}
f=\frac{1}{2}|v|^2+\frac{\lambda}{2}|\nabla d|^2+\lambda \varPhi(d)-s(C,g) ,
\end{equation}
where $\lambda>0$ stood for the ratio between kinetic energy and potential energy. $\varPhi(d)=\frac{1}{2\epsilon^2}(|d|^2 -1 )^2$ was a penalty function, whose derivative was given by
$\varphi(d)=\frac{\partial \varPhi}{\partial d}=\frac{1}{\epsilon^2}(|d|^2 -1 )d$. It was direct to see that,
$\varphi(d)$ was the Ginzburg-Landau approximation of the constraint that the director had a unit length $|d|=1$, when $\epsilon$ was small \cite{Lin2000Existence}.
Moreover, $s(C,g)$ was the entropy function of non-equilibrium state variables $(C,g)$, which were conjugate variables of $(\tau_v, q)$ with respect to free energy $f$, that is, $f_{C}=\tau_v, f_{g}=q$. Again, the temperature $T$ was set to be one for simplicity.

Then, according to the generalized Gibbs relation, the time evolution of the free energy $f=f(v,d,\nabla d, C,g)$ was calculated as
\begin{align*}
    \frac{df}{dt}
    =&v\cdot \frac{dv}{dt}+\lambda \nabla d : (\frac{d}{dt}\nabla d)+\lambda \varphi(d)\cdot \frac{d}{dt} d + f_{C}: \frac{dC}{dt} + f_{g}\cdot \frac{dg}{dt}\\
    =&v\cdot \frac{dv}{dt}+\lambda \nabla d : [ \nabla (\frac{d}{dt} d) - \nabla v \cdot \nabla d] + \lambda \varphi(d)\cdot \frac{d}{dt} d+ f_{C} : \frac{dC}{dt} + f_{g}\cdot \frac{dg}{dt}\\
    =&v\cdot (\nabla \cdot \tau)+\lambda \nabla d : [ \nabla q - \nabla v \cdot \nabla d] + \lambda \varphi(d)\cdot q+  \tau_v : \frac{dC}{dt} + q \cdot \frac{dg}{dt}\\
    =&\nabla \cdot [\tau\cdot v + \lambda (\nabla d) \cdot q]- \tau:\nabla v  - \lambda \Delta d \cdot q- \lambda \nabla d :(\nabla v \cdot \nabla d)+ \lambda \varphi(d)\cdot q \\
     &  + \tau_v : \frac{dC}{dt} + q \cdot \frac{dg}{dt}\\
   =&\nabla \cdot J^f + q \cdot [ \frac{dg}{dt} + \lambda( \varphi(d) -  \Delta d) ] + \tau_v : (\frac{dC}{dt}- \nabla v)
     - \nabla v :[\tau_e +  \lambda \nabla d \cdot (\nabla d)^T ]
   ,
\end{align*}
where the entropy flux $J^f=\tau\cdot v + \lambda (\nabla d) \cdot q$.
During the derivation, we have used the relation $\frac{d}{dt}\nabla d=\nabla (\frac{d}{dt} d) - \nabla v \cdot \nabla d$ in the second step and the identity $(\nabla d)^T:(\nabla v \cdot \nabla d) = \nabla v: [\nabla d \cdot (\nabla d)^T]$ in the last step.

In accordance with the second law of thermodynamics, we concluded that the entropy production rate
\begin{equation}
\sigma^f=-q \cdot [ \frac{dg}{dt} + \lambda( \varphi(d) -  \Delta d) ] - \tau_v : ( \frac{d C}{dt}- \nabla v)
+ \nabla v :[\tau_e +  \lambda \nabla d \cdot (\nabla d)^T ]
\geq 0\nonumber
\end{equation}
was non-negative.
CDF suggested following constitutive equations for $\tau_e, \tau_v$ and $q$:
\begin{align}
&\tau_e(d)=-  \lambda \nabla d \cdot (\nabla d)^T,\\
& \frac{d C}{dt}- \nabla v= -\frac{1}{\alpha} \tau_v, \label{constitutive1}\\
& \frac{d g}{dt} + \lambda( \varphi(d) -  \Delta d)= -\frac{1}{\beta} q, \label{constitutive2}
\end{align}
where the elastic stress
$\tau_e$
had no contribution to the entropy production rate.
Notice that the dissipation matrix adopted in Eqs. \eqref{constitutive1}-\eqref{constitutive2} was a diagonal one,
with $\alpha, \beta>0$ representing typical relaxation time for the viscous stress and force moment respectively.

Finally, we specified the entropy function $s(C,g)$ as
\begin{equation}
s=-\frac{1}{2\gamma}|C|^2 - \frac{1}{2\gamma}|g|^2,
\end{equation}
where the coefficient $\gamma>0$.
Then $C=\gamma \tau_v, ~g=\gamma q$ by definition. Applying the Maxwell iteration on Eqs. \eqref{constitutive1}-\eqref{constitutive2} in the limit of $\alpha, \beta \rightarrow 0$, and then substituting them into Eq. \eqref{liquid-1}, we arrived at the simplified E-L equations for hydrodynamic flows of nematic liquid crystals proposed by Lin and Liu \cite{Lin1995Nonparabolic}. That is
\begin{equation}\label{liquid-2}
\begin{aligned}
&\nabla \cdot v=0,\\
&\frac{\partial v}{\partial t}+(v \cdot \nabla v)=-\nabla p +\nabla \cdot [\alpha \nabla v -\lambda \nabla d \cdot (\nabla d)^T],\\	
&\frac{\partial d}{\partial t}+(v \cdot \nabla d) = -\beta \lambda( \varphi(d) -  \Delta d).
\end{aligned}
\end{equation}

Notice that, the equation for angular momentum in Ref. \cite{Liu2009On} had an additional term $-(\nabla v)^T\cdot d $,
the upper-convected time derivative, to fulfill the principle of material frame indifference.
Taking this term into account and repeating the same procedure listed above,
we could also recover the hydrodynamical model for nematic liquid crystals given in Ref. \cite{Liu2009On}, except that
the resulting elastic stress became
$\tau_e=-  \lambda \nabla d \cdot (\nabla d)^T+ \lambda (\varphi(d)-\nabla d) \otimes d$.

\section{Conclusions and Discussions}
In this work, we have shown that the variational approach proposed by M. Doi was equivalent to CDF by choosing the proper variational function and variables.
The correspondence between two theories has been further illustrated through several novel examples in soft matter physics, including particle diffusion in dilute solutions, polymer phase separation dynamics and hydrodynamic flows of liquid crystals in the nematic phase. Our results not only validated the usefulness of CDF, which has been used in the current case to regularize the variational approach and put it on a more rigorous mathematical foundation, but also provided a great convenience for future studies on the modeling of various non-equilibrium processes, since either CDF or the variational approach could be adopted with the same outcome.

It was well known that, in response to surrounding hydrodynamic flows, polymers would change their conformations from time to time, which was generally characterized through the configuration tensor (or $Q$-tensor). The mathematical theory for polymeric fluids (or complex fluids) by using $Q$-tensor was started from Kirkwood in the 1940s \cite{Kirkwood1946The,Kirkwood1947The}, then followed by Bird \textit{et al.} \cite{Bird1987Dynamics}, Doi and Edwards \cite{doi1988theory}, and many others. The scalar model we considered in Section \ref{example2} could be regarded as a simplified version of the $Q$-tensor theory. In fact, it was shown that the Ericksen-Leslie theory could be recovered from the $Q$-tensor model by making uniaxial assumptions \cite{Han2015From}. Lin and Liu \cite{Lin1995Nonparabolic} further simplified the E-L theory and deduced the scalar model we used, in which many mathematical properties of interest of the original model were preserved.

\section*{acknowledgment}
This work was supported by the 13th 5-Year Basic Research Program of CNPC (2018A-3306), the National Natural Science
Foundation of China (Grants 21877070) and Tsinghua University Initiative Scientific Research Program (Grants 20151080424). The authors would like to thank the helpful discussions from Dr. Zhiting Ma and Xiaokai Huo.

\bibliographystyle{ieeetr}
\bibliography{ref}
\end{document}